\documentclass[twocolumn]{aa}
\usepackage{graphicx}
\usepackage{txfonts}
\def\kms{\relax \ifmmode {\,\rm km\,s}^{-1}\else \,km\,s$^{-1}$\fi}
\def\ks{\relax \ifmmode  K_{\rm s}\else $K_{\rm s}$\fi}
\def\ha{\relax \ifmmode {\rm H}\alpha\else H$\alpha$\fi}
\def\hb{\relax \ifmmode {\rm H}\beta\else H$\beta$\fi}
\def\hi{\relax \ifmmode {\rm H\,{\sc i}}\else H\,{\sc i}\fi}
\def\hii{\relax \ifmmode {\rm H\,{\sc ii}}\else H\,{\sc ii}\fi}
\def\h2{\relax \ifmmode {\rm H}_2\else H$_2$\fi}
\def\lha{\relax \ifmmode L_{{\rm H}\alpha}\else $L_{{\rm H}\alpha}$\fi}
\def\shi{\relax \ifmmode \sigma_{{\rm HI}}\else $\sigma_{\rm HI}$\fi}
\def\sh2{\relax \ifmmode \sigma_{{\rm H}_2}\else $\sigma_{{\rm H}_2}$\fi}
\def\degr{\hbox{$^\circ$}}
\def\arcmin{\hbox{$^\prime$}}
\def\arcsec{\hbox{$^{\prime\prime}$}}
\def\deg{\hbox{$^\circ$}}

\def\fdg{\hbox{$.\!\!^\circ$}}
\def\fs{\hbox{$.\!\!^{\rm s}$}}
\def\farcm{\hbox{$.\mkern-4mu^\prime$}}
\def\farcs{\hbox{$.\!\!^{\prime\prime}$}}
\def\degd#1.#2{ #1\fdg#2 }                 
\def\mind#1.#2{ #1\farcm#2 }               
\def\secd#1.#2{ #1\farcs#2 }               
\def\hhh{\ifmmode {\rm ^h}              
         \else {${\rm ^h}$}
         \fi}
\def\sss{\ifmmode {\rm ^s}              
         \else {${\rm ^s}$}
         \fi}
\def\hms#1h#2m#3s{                      
                  \relax
                  \ifmmode #1^{\rm h}\,#2^{\rm m}\,#3^{\rm s}
                  \else \hbox{$#1^{\rm h}\,#2^{\rm m}\,#3^{\rm s}$}
                  \fi
                 }
\def\dms#1d#2m#3s{                      
                  \relax
                  #1\degr\,#2\arcmin\,#3\arcsec 
                 }
\def\hmsd#1h#2m#3.#4s{                  
                      \relax
                      \ifmmode #1^{\rm h}\,#2^{\rm m}\,#3\fs#4
                      \else \hbox{$#1^{\rm h}\,#2^{\rm m}\,#3\fs#4$}
                      \fi
                     }
\def\dmsd#1d#2m#3.#4s{                  
                      \relax
                      #1\degr\,#2\arcmin\,#3\farcs#4
                     }
\def\mag{\relax                          
        \ifmmode ^{\rm m}
        \else $^{\rm m}$
        \fi
       }
\def\magd#1.#2{                          
              \relax
              \ifmmode #1^{\rm m}
                       \hskip-0.55em.\hskip0.22em#2
              \else \hbox{#1$^{\rm m}
                    \hskip-0.55em.\hskip0.22em$#2}
              \fi
             }
\input psfig.sty
\begin{document}
\title{Structure and star formation in disk galaxies}
\subtitle{II. Optical imaging}
\author{J.~H.~Knapen\inst{1}
\and S.~Stedman\inst{1}
\and D.~M.~Bramich\inst{2}
\and S.~L.~Folkes\inst{1}
\and T.~R.~Bradley\inst{1}
}

\offprints{J. H. Knapen}  

\institute{Centre for Astrophysics Research, University of
Hertfordshire, Hatfield, Herts AL10 9AB, U.K.\\
\email{j.knapen@star.herts.ac.uk} 
\and School of Physics and Astronomy, University of St. Andrews,
Scotland KY16 9SS\\}

\date{Received ; accepted 20/07/2004}

\abstract{ We present optical observations of a sample of 57 spiral
galaxies and describe the procedures followed to reduce the data.  We
have obtained images in the optical $B$ and $I$ broad bands, as well
as in \ha, with moderate spatial resolution and across wide enough
fields to image the complete disks of the galaxies.  In addition, we
observed 55 of our sample galaxies in the $R$ and eight in the $V$
band, and imaged a subset through a dedicated narrow continuum filter
for the \ha\ line.  We describe the data reduction procedures we
developed in the course of this work to register, combine and
photometrically calibrate each set of images for an individual galaxy.
We describe in some detail the procedure used to subtract the
continuum emission from our \ha\ images.  In companion papers, we
describe near-infrared imaging of the galaxy sample, and present
analyses of disk scale lengths, and of properties of bars, rings, and
\hii\ regions in the sample galaxies.  The images described here are
available for use by other researchers.


}

\maketitle

%

\section{Introduction}

The structure of disk galaxies has been studied in a quantitative way
for decades now (e.g., de Vaucouleurs 1948; Freeman 1970) and although
the field has much advanced since the early days of photographic
plates, several old problems have not yet been solved, and new
problems have appeared.  Among the most important of such problems are
the nature of the exponential disk, the evolution of bulges, the
origin of spiral structure and its influence on star formation, the
life cycle of bars, and the origin and influence of dark matter.
Imaging of disk galaxies has evolved from the early days, when Hubble
(e.g., 1926) used the largest telescopes available at the time to
photograph galaxies, to fully automated all-sky surveys in the near-IR
(e.g., 2MASS: Jarrett et al. 2003).

In planning the programme of which the results described here form
part, we identified the need for a technically uniform set of images
covering the complete disks of a well-defined sample of spiral
galaxies. One of the main aims of this overall programme is the study
of the distribution of scale lengths in different classes of spiral
galaxies, and specifically to study possible differences between arm
and interarm regions.  This drove us to develop a significant imaging
programme, of near-IR \ks-band imaging on the one hand (described by
Knapen et al.  2003, hereafter Paper~I), and more conventional optical
broad- and \ha\ narrow-band imaging on the other (this paper).  Our
sample consists of 57 nearby and not highly inclined spiral galaxies
of all types, and since it contains a large number of galaxies which
continue to attract significant attention from researchers (e.g., M74,
M77, M95, M100) the importance of the currently presented set of
images, along with those of Paper~I, goes beyond our immediate
research aims, which will be explored in a number of subsequent
papers.  The combination of these factors justifies the current
publication of the details of the data collection and reduction.
Workers in the field are welcome to peruse our images, through the
CDS\footnote{Images are available in electronic form at the CDS via
anonymous ftp to cdsarc.u-strasbg.fr (130.79.128.5) or via
http://cdsweb.u-strasbg.fr/cgi-bin/qcat?J/A+A/}, and/or reduction
scripts.

After briefly summarising the sample selection in Sect.~2 of this
paper, we describe the observations and the reduction of the data in
Sects.~3 and~4, respectively. The description of the final data set,
in Sect.~5, is followed by a short concluding section (Sect.~6).

\section{Sample}

The selection criteria for the sample observed, as well as some of its
statistical properties, have been described in detail in Paper~I, and
we only summarise the main points here.  We selected all galaxies
which are larger than 4.2~arcmin in diameter, of spiral type, inclined
less than 50\deg, and visible from the Northern hemisphere
($\delta>-20$\deg).  The resulting sample contains spiral galaxies of
all morphological spiral types, with and without bars, with and
without nuclear activity of Seyfert, LINER, or starburst type, and of
all spiral arm classes (cf.  Elmegreen \& Elmegreen 1987).  All sample
galaxies are relatively nearby, with an upper limit in systemic
velocity of close to 2500\,\kms.  Apart from the optical observations
described in the present paper, we also obtained images of all sample
galaxies in the near-IR \ks-band. As described in Paper~I, the latter
images cover the complete disk of the galaxy in most cases, and most
of the disk for all galaxies, and are almost all of sub-arcsec spatial
resolution.

\section{Observations}

\begin{table*}
\hfill
\centering
\begin{tabular}{llcccccccccccc}
\hline
\multicolumn{2}{c}{Galaxy} & \multicolumn{4}{c}{Exposure
time (min.)} & \multicolumn{4}{c}{Seeing (arcsec)} & H$\alpha$ filter & Cont. filter &
Scale & Image size\\
NGC & Messier & $B$ & $R$ & $I$ & H$\alpha$ & $B$ & $R$ & $I$ &
H$\alpha$ & $\lambda /\Delta\lambda$ & & factor & (arcmin)\\
\hline
210 & & 20 & 15 & 40 & 40 & 1.6 & 1.6 & 1.4 & 1.3 & 6594/44 & 6470/115 & 0.516 & 8.1 \\
337A & & 50 & 10 & 30 & 60 & 1.9 & 1.4 & 1.3 & 1.3 & 6594/44 & $R$ & 0.026 & 8.1 \\
488 & & 35 & 10 & 10 & 60 & 1.5 & 1.5 & 1.9 & 1.5 & 6607/50 & $R$ & 0.031 & 8.1 \\
628 & 74 & 36 & 30 & 20 & 40 & 1.4 & 1.8 & 1.2 & 1.0 & 6570/55 & $I$ & N/A & $11.2\times22.6^2$ \\
864 & & 35 & 30 & 35 & 120 & 1.6 & 1.1 & 1.6 & 1.4 & 6594/44 & $R$ & 0.062 & 6.9 \\
1042 & & 20 & 10 & 25 & 90 & 1.4 & 1.7 & 1.4 & 1.5 & 6594/44 & $R$ & 0.021 & 8.1 \\
1068 & 77 & 35 & 10 & 30 & 40 & 2.4 & 2.2 & 1.4 & 2.0 & 6594/44 & $R$ & 0.019 & 9.5 \\
1073 & & 20 & 15 & 10 & 60 & 3.7 & 1.2 & 1.3 & 1.4 & 6594/44 & $R$ & 0.018 & 8.9 \\
1169 & & 30 & 15 & 20 & 60 & 1.4 & 1.2 & 1.2 & 1.8 & 6626/44 & $R$ & 0.019 & 8.1 \\
1179 & & 20 & 10 & 20 & 80 & 2.1 & 2.2 & 1.8 & 1.9 & 6594/44 & $R$ & 0.018 & 6.5 \\
1300 & & 30 & -- & 30 & 40 & 2.7 & -- & 2.3 & 1.6 & 6594/44 & $I$ & 0.012 & 6.1 \\
2775 & & 15 & 15 & 10 & 100 & 1.4 & 1.8 & 1.4 & 1.9 & 6594/44 & $R$ & 0.044 & 5.6 \\
2805 & & 15 & 15 & 20 & 120 & 1.3 & 1.4 & 1.1 & 1.7 & 6594/44 & $R$ & 0.048 & 6.5 \\
2985 & & 45 & 15 & 10 & 60 & 1.8 & 1.4 & 1.8 & 1.8 & 6594/44 & $R$ & 0.045 & 6.1 \\
3184 & & 30 & 12 & 40 & 80 & 1.7 & 1.7 & 1.4 & 1.7 & 6570/55 & $R$ & 0.024 & 8.1 \\
3227 & & 60 & 15 & 20 & 80 & 1.9 & 1.4 & 1.7 & 1.2 & 6594/44 & $R$ & 0.020 & 7.3 \\
3344 & & 15 & 15 & 10 & 20 & 1.2 & 1.4 & 1.4 & 1.1 & 6570/55 & $R$ & 0.025 & 8.9 \\
3351 & 95 & 30 & 30 & 10 & 60 & 1.7 & 1.9 & 1.6 & 1.6 & 6570/55 & 6470/115 & 0.500 & 6.5 \\
3368 & 96 & 60 & 15 & 20 & 20 & 2.3 & 1.1 & 1.6 & 1.7 & 6594/44 & $R$ & 0.054 & 9.3 \\
3486 & & 15 & 15 & 20 & 60 & 1.2 & 1.3 & 1.3 & 1.3 & 6570/55 & $R$ & 0.057 & 6.05 \\
3631 & & 45 & 10 & 30 & 30 & 1.6 & 2.2 & 1.3 & 1.3 & 6589/15 & 6565/15 & N/A$^1$ & 5.6 \\
3726 & & 15 & 15 & 30 & 60 & 1.2 & 1.4 & 1.5 & 1.3 & 6570/44 & $R$ & 0.023 & 6.5 \\
3810 & & 15 & 10 & 10 & 100 & 1.2 & 1.4 & 1.2 & 1.3 & 6594/44 & 6470/115 & 0.425 & 7.3 \\
4030 & & 30 & 15 & 10 & 60 & 1.7 & 1.6 & 1.8 & 1.5 & 6594/44 & $R$ & 0.024 & 6.9 \\
4051 & & 30 & 10 & 20 & 60 & 1.4 & 1.7 & 1.4 & 1.5 & 6570/55 & 6470/115 & 0.436 & 7.7 \\
4123 & & 30 & 15 & 20 & 60 & 2.3 & 1.3 & 1.4 & 1.3 & 6594/44 & $R$ & 0.020 & 7.7 \\
4145 & & 15 & 20 & 20 & 40 & 2.5 & 1.4 & 2.5 & 2.7 & 6594/44 & $R$ & 0.025 & 6.1 \\
4151 & & 15 & 15 & 10 & 60 & 1.9 & 1.5 & 1.4 & 1.4 & 6594/44 & $R$ & 0.022 & 6.9 \\
4242 & & 15 & 15 & 10 & 60 & 1.4 & 1.2 & 1.6 & 3.2 & 6570/55 & $R$ & 0.050 & 8.7 \\
4254 & 99 & 15 & 15 & 10 & 40 & 2.1 & 1.1 & 1.2 & 1.5 & 6626/44 & $R$ & 0.055 & 10.2 \\
4303 & 61 & 30 & 15 & 50 & 60 & 1.6 & 1.3 & 1.7 & 1.4 & 6594/44 & $R$ & 0.042 & 8.5 \\
4314 & & 15 & 15 & 10 & 20 & 1.4 & 1.8 & 1.3 & 1.3 & 6594/44 & $R$ & 0.240 & 7.8 \\
4321 & 100 & 10 & 5 & 5 & 20 & 1.9 & 2.1 & 1.9 & 1.0 & 6601/15 & 6565/15,6577/15 & N/A$^1$ & 10.1 \\
4395 & & 15 & 20 & 9 & 30 & 1.9 & 1.2 & 1.6 & 1.8 & 6568/95 & $R$ & 0.660 & 11.4 \\
4450 & & 15 & 15 & 10 & 60 & 1.6 & 0.8 & 1.5 & 0.9 & 6607/50 & $R$ & 0.025 & 6.5 \\
4487 & & 15 & 15 & 10 & 60 & 3.2 & 1.2 & 2.9 & 2.0 & 6594/44 & $R$ & 0.047 & 6.2 \\
4535 & & 30 & 15 & 10 & 60 & 1.9 & 1.6 & 1.7 & 1.9 & 6594/44 & $R$ & 0.042 & 9.9 \\
4548 & 91 & 30 & 10 & 20 & 60 & 1.5 & 1.7 & 1.4 & 1.7 & 6594/44 & 6470/115 & 0.418 & 8.2 \\
4579 & 58 & 30 & 15 & 20 & 60 & 1.6 & 1.1 & 1.8 & 1.6 & 6594/44 & $R$ & 0.050 & 8.6 \\
4618 & & 15 & 15 & 10 & 60 & 1.5 & 0.8 & 1.5 & 1.3 & 6570/55 & $R$ & 0.025 & 7.9 \\
4689 & & 45 & 15 & 10 & 60 & 1.9 & 0.8 & 1.5 & 1.8 & 6594/44 & $R$ & 0.038 & 8.6 \\
4725 & & 30 & 15 & 10 & 40 & 2.2 & 1.2 & 2.1 & 2.3 & 6594/44 & $R$ & 0.038 & 11.4 \\
4736 & 94 & 10 & 27 & 9 & 60 & 1.9 & 1.3 & 2.2 & 1.4 & 6570/55 & $R$ & 0.043 & 11.4 \\
5247 & & 15 & 15 & 10 & 60 & 2.6 & 2.3 & 2.4 & 2.3 & 6594/44 & $R$ & 0.018 & 8.5 \\
5248 & & 15 & 30 & 10 & 60 & 2.2 & 1.3 & 2.0 & 1.4 & 6594/44 & $R$ & 0.005 & 8.5 \\
5334 & & 15 & 15 & 30 & 60 & 1.6 & 1.4 & 2.3 & 1.8 & 6594/44 & $R$ & 0.020 & 8.5 \\
5371 & & 15 & 15 & 10 & 60 & 1.4 & 1.1 & 1.2 & 1.3 & 6626/44 & $R$ & 0.040 & 8.1 \\
5457 & 101 & 15 & 15 & 9 & 60 & 3.0 & 1.3 & 2.3 & 2.4 & 6570/55 & $R$ & 0.250 & 11.5 $\times$ 22.9$^2$ \\
5474 & & 15 & 15 & 10 & 60 & 1.3 & 1.2 & 1.3 & 1.2 & 6570/55 & $R$ & 0.047 & 8.4 \\
5850 & & 15 & 15 & 10 & 60 & 1.4 & 1.6 & 1.3 & 1.5 & 6626/44 & $R$ & 0.019 & 8.5 \\
5921 & & 15 & 15 & 10 & 40 & 1.4 & 1.3 & 1.4 & 1.2 & 6594/44 & $R$ & 0.088 & 8.5 \\
5964 & & 15 & 15 & 20 & 80 & 1.8 & 1.3 & 1.3 & 1.8 & 6594/44 & $R$ & 0.048 & 8.5 \\
6140 & & 45 & 10 & 30 & 60 & 1.5 & 2.0 & 1.2 & 1.4 & 6594/44 & $I$ & 0.020 & 7.9 \\
6384 & & 30 & 15 & 40 & 60 & 1.7 & 1.7 & 1.4 & 1.8 & 6594/44 & $R$ & 0.040 & 8.1 \\
6946 & & 15 & 15 & 20 & 60 & 1.4 & 1.5 & 1.2 & 1.4 & 6570/55 & $R$ & 0.045 & 11.5 \\
7727 & & 15 & -- & 30 & 60 & 1.7 & -- & 1.4 & 1.7 & 6594/44 & 6470/115 & 0.520 & 7.3 \\
7741 & & 15 & 10 & 30 & 80 & 1.6 & 1.4 & 1.5 & 1.5 & 6570/55 & $R$ & 0.023 & 8.1 \\
\hline
\end{tabular}

\caption{Details of the images obtained for our sample galaxies: total
on-source exposure time in minutes (col.~3-6); seeing in arcsec as
measured on the reduced images (col.~7-10); filter used for the \ha\
line, where $\lambda /\Delta\lambda$ is the central wavelength and
filter width, in \AA\, (col.~11) and continuum (col.~12) imaging;
scale factor used to subtract the continuum from \ha\ (col.~13); and
size of the final image set in arcmin (col.~14).  Where only one
number is given the image is approximately square. Notes: $^1$ see
Sect. 4.4; $^2$ $B$-band and \ha\ images are only $8.0\times8.0$
arcmin.}

\label{imagetab}
\end{table*}

We present a complete set of images in the broad Harris $B$ and $I$
bands and in the narrow \ha\ band of our sample of 57 galaxies, and in
addition $R$-band images of 55 of our 57 sample galaxies.  Most of
these images were obtained on the 1-m Jacobus Kapteyn Telescope (JKT)
on La Palma, during a number of observing runs from 1999-2003. We also
obtained images from the JKT through the service programme, where
observatory staff performed the observations.  A number of images,
especially of the larger galaxies in our sample, were obtained with
the 2.5-m Isaac Newton Telescope (INT) on La Palma ($B$: NGC~628,
NGC~1073, NGC~3368, NGC~4254, NGC~4321, NGC~4395, NGC~4535, NGC~4725,
NGC~4736; $R$: NGC~4321, NGC~5248; $I$: NGC~628, NGC~4321, NGC~4395,
NGC~4535, NGC~4725, NGC~4736, 5457; \ha: NGC~4314, NGC~4395,
NGC~5457). In addition to these new observations, we also made
extensive use of the Isaac Newton Group (ING) data archive.  After
discarding a significant fraction of the images we originally found in
the archive, about 15\% of our final images were obtained from there.
Reasons for discarding images, and repeating the observations, range
from too old observations with too small a field of view (FOV), to
lack of documentation on, e.g., the filters used.  For NGC~3631 and
NGC~4321, we used \ha\ images from the literature (see Sect.~4.4),
while the broad-band images of NGC~4321 date back to Knapen et
al. (1993a). The $R$-band image of NGC~5248 has been described in
detail by Jogee et al. (2002).

A considerable variety of cameras and CCD detectors was used, but the
principal ones were the camera on the JKT, imaging onto a SITe2 CCD
with 2048$\times$2048 pixels of a projected size of 0.331~arcsec each,
and the wide field camera (WFC) on the INT. The latter has four EEV
CCDs of 2048$\times$4096 pixels, also with a projected pixel size of
0.331~arcsec.  The unvignetted FOV of the JKT camera is
$\sim$10~arcmin, that of each of the four WFC CCDs is 11\,$\times$\,22
arcmin.  We used a set of \ha\ filters with central wavelengths
matched to the recession velocities of the galaxies, and width in most
cases around 50~\AA, as listed in Table~\ref{imagetab}.  Seeing
conditions varied during the course of our observational campaigns and
the seeing as measured from the final images ranges from 0.8 to
3.7~arcsec, with a median of 1.64, 1.39, 1.45 and 1.53 arcsec FWHM for
the $B$, $R$, $I$ and \ha\ images, respectively. Seeing values for the
individual images are listed in Table~\ref{imagetab}.

All observations were accompanied by a sufficient number of bias and
dusk or dawn sky flat fields for calibration (lack of such exposures
was one possible reason for discarding archive images).  On
photometric nights, a number of Landolt (1992) standard stars and
spectrophotometric standards were observed in order to provide the
necessary photometric zero points (see Sect.~4.5).

\section{Data reduction}

The main problem we faced during the reduction of the large amounts of
data we collected was the combination of images taken on different
nights, under different conditions, and often with different cameras,
pixel sizes, FOV, or orientation of the images on the sky. To manage
this problem, we wrote a set of scripts in IRAF\footnote{IRAF is
distributed by the National Optical Astronomy Observatories, which is
operated by the Association of Universities for Research in Astronomy,
Inc.  (AURA) under cooperative agreement with the National Science
Foundation.}, aimed to handle the complete data reduction, including
bias subtraction, flatfielding, and sky subtraction.  We use a
combination of IDL and IRAF scripts to subtract a scaled continuum
from the \ha\ images.  Each interactive script handles a specific
aspect of the data reduction pipeline providing the advantages of
increased efficiency and data quality control at each step without
committing to the whole data reduction process in one go.  We will
describe some of these scripts in more detail below, but give a
complete list with short descriptions here. 

\begin{itemize}

\item[--] {\tt bias.cl} - Calculate the mean bias level as a constant
from a set of bias frames.

\item[--] {\tt flat.cl} - Create a mean flat field frame from a set of
flat field frames and using a constant as the bias level.

\item[--] {\tt reduce.cl} - Subtract the bias level as a constant from a
set of science frames and flat field them all using a specified flat
field frame.

\item[--] {\tt meanimcombine.cl} - Combine a set of images of one
object in a single waveband by averaging the images weighted by their
exposure times, and performing the required scaling, rotation, and
translation. The output image is scaled to an exposure time of one
second and an option exists to align it to a Digitized Sky Survey
(DSS) image.

\item[--] {\tt medianimcombine.cl} - As {\tt meanimcombine.cl} but
using the weighted median rather than the weighted mean of the input
images.

\item[--] {\tt background.cl} - Subtract a constant  sky background 
level from the image after determining  the correct value by iteratively
averaging and 3 sigma clipping across a user-defined area of background.

\item[--] {\tt lineup.cl} - Lines up images of the same galaxy in
different wavebands, scaling the  images and changing their canvas  size
so that no data are lost in the process.

\item[--] {\tt aminusb.cl} - Creates an $A-B$ image which maps the
colour differences across the image, in magnitudes. The output image
is produced after smoothing the image with the best seeing to match
the seeing of the other by convolving it with the required normalised
2D Gaussian function.
 
\item[--] {\tt smooth.cl} - Smooths image with the best seeing to match
the seeing of the other by convolving it with the required normalised 2D
Gaussian function.

\item[--] {\tt contsub.pro} - An IDL program that aids in determining
the correct scaling factor for the continuum images for producing the
continuum subtracted \ha\ maps by plotting the continuum counts versus
the \ha\ counts for each pixel and measuring the gradient.

\end{itemize}

\subsection{Bias subtraction and flat fielding}

All bias and sky flat field frames were inspected for quality and
those with very low or high (non-linear or saturated) counts, or with
abnormal structure or read out noise, were removed.  Since for all the
CCDs used for this project the structure in the bias frames is
negligible, we subtracted a constant bias level from all images. The
SITe2 chip on the JKT has a typical bias level of $\sim$625~ADU in the
``quick" mode that was used, and a read out noise of $\sim$5.3~ADU.
With an average of nine bias frames taken per night, this leads to an
uncertainty of $\sim$2~ADU, or $\sim$0.3\%, in the bias level.

After checking the internal consistency of the flat field frames for
each filter and each night, they were entered into the {\tt flat.cl}
script, which we used to create a normalised master flat, one per
night per filter.  Assuming typical numbers for bias level and gain,
the uncertainty in a typical normalised master flat is of order of
0.35\%.

We next used the script {\tt reduce.cl} to bias-subtract and flat
field all images taken through a certain filter in a certain night.
After removing the outer areas of the chip, the reduced science frames
are passed once through the {\sc cosmicrays} task in IRAF to remove in
an automated way the majority of hot pixel events.

\subsection{Image combination}

At this point in the data reduction procedure there is usually more
than one image per filter per object.  This is an advantage because if
these images are of comparable quality they may be combined to produce
an image with a greater signal to noise ratio.  The images may be
oriented differently, however, have a different FOV, and may have been
obtained with different instruments, and under different conditions.
After renaming the multiple individual images appropriately, they are
entered into the script {\tt meanimcombine.cl}, along with a DSS image
if required.  The script does the following:

\begin{itemize}

\item[1] Divide each image by its exposure time in seconds, as
obtained from the header.

\item[2] Ask the user to mark interactively three stars common to all
the images (and two of these common to the DSS image).

\item[3] Map each image onto the image with the smallest
pixels (which is effectively the most detailed image) while preserving
flux, and using parameters for geometric scaling, rotation and
translation as determined from the fits to the star positions.  This
can lead to some loss of information if the FOV of the most detailed
image is smaller than that of one or more of the other images, an
effect that we prevent by manually copying the former image onto a
larger pixel grid.  This script may also lead to blank areas on some
images where data does not exist.

\item[4] Combine the images by averaging the pixel values, weighted by
the exposure times of the images, and ignoring blank pixels, to
produce a combined output image.  We have a second version of the
script ({\tt medianimcombine.cl}) which combines images by using the
median value of the pixels, in the process removing most of the cosmic
ray hits or otherwise bad pixels. This latter script is used only on
images which are virtually identical, because for instance a different
background or higher counts level in one of three images can lead to
an unacceptable loss of information when median combination is used.

\item[5] Using the two common stars marked in the DSS image, the
script rotates the output image in order to line it up to the DSS
image. This results in an output image where North is up, and East to
the left, to an accuracy of better than 1\deg.

\end{itemize}

The combined output image has an effective exposure time of 1~sec
which is represented in its header by the keyword EXPTIME=1.  The
image represents a total exposure time, however, equal to the sum of
all the individual exposure times of the images used in the
combination algorithm. We include this total exposure time, in
minutes, in the file name of the combined image, as well as in the
header of the FITS file. At this point, there is only one image per
filter per object.

\subsection{Sky background subtraction and image registration}

The script {\tt background.cl} allows the user to mark interactively
an area of approximately flat sky background, ideally far removed from
the galaxy being imaged. Since this area may well include a couple of
stars or cosmic rays, the script incorporates an interactive $3\sigma$
clip algorithm with a tolerance of 0.02\% to derive the underlying
background level. This level is then subtracted as a constant from the
whole image.  The method works well when the image has a flat
background, and large areas without galaxy emission.  It delivers
merely an approximation, however, when the image has no areas free of
galaxy emission, or when it shows a background with structure, e.g.,
due to non-perfect flat fielding.  In these cases, the script {\tt
background.cl} was used to estimate the background level by running it
on the flattest areas with the lowest counts available in the image.

To register the sets of images associated with a particular galaxy,
all the images in that set (at least $B$, $I$, \ks, and \ha) are
entered into the script {\tt lineup.cl}.  This script uses part of the
functionality of the {\tt meanimcombine.cl} script (see above), and
again requires a common set of three stars to be marked in each
image. The script identifies the image with the smallest pixels (i.e.,
the most detailed image) and re-grids all other images so these will
have the same pixel scale as the most detailed image. The script then
maps all images onto the image with the largest FOV, thus ensuring
that no information is lost.  Any areas in the images that lack data
are marked as blank by setting these pixels to zero.  This approach
has a drawback which is that the size of the image files grows due to
both smaller pixels and larger areas. By compressing the images much
can be gained though, because the areas on the outside of the images
which are filled by pixels with value zero can be very effectively
compressed.  Flux is conserved throughout this procedure.  Finally,
the images are trimmed to a standard and usually square size.

\subsection{Continuum subtraction from the \ha\ images}

In order to subtract the continuum emission from our \ha\ images, two
different techniques were used, depending on the origin and quality of
the images and on the kind of continuum image available. For NGC~3631
and NGC~4321 we used \ha\ images obtained with the WHT, for which the
continuum subtraction technique has been described in detail by Knapen
et al.  (1993b), with more specific details given by Rozas, Beckman \&
Knapen (1996) for NGC~3631 and by Knapen (1998) for NGC~4321. As
detailed in Table~\ref{imagetab}, for most of the remaining galaxies
we used a broad-band $R$ image, whereas for the others we used either
an image obtained through a special \ha\ continuum filter with central
wavelength 6470\AA\ and FWHM 110\AA (hereafter called 6470/110), or an
$I$-band image.

\begin{figure}
\centering
\psfig{figure=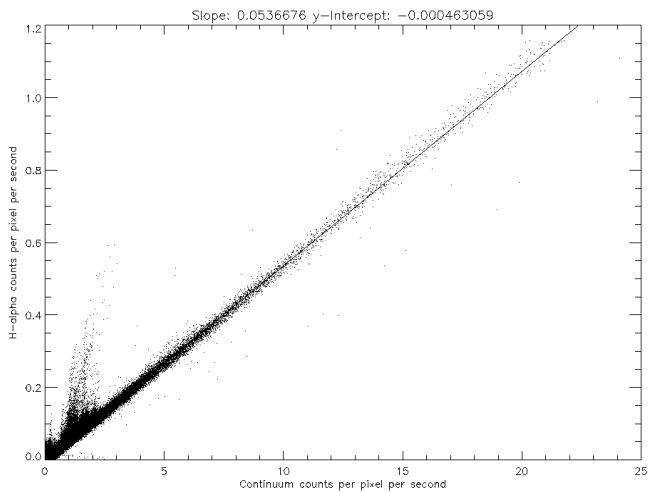,angle=0,width=9cm}
\caption{Intensity, in units of counts per pixel per second, of each
pixel in the \ha\ image of NGC~3368 versus the intensity of the same
pixel in the continuum image, taken through the 6470/115 filter.  Most
points outline continuum-only emission, and the drawn line is a fit
which gives an estimate of the scaling factor to be applied to the
continuum image before subtracting it from the \ha\ image. A limited
number of pixels traces \ha\ emission in the galaxy, and can be seen
above the continuum band.}
\label{fig1}
\end{figure}

Since at this stage in the reduction the images have been aligned,
registered and background-subtracted, the one remaining parameter to
be determined is the scaling factor, which determines how the
continuum image must be scaled to match the intensity level of the
continuum emission in the \ha\ line image.  For each galaxy, we
smoothed the image with the highest resolution to that of the other.
We then used a combination of two techniques to derive the scaling
factor.  In the first, we basically follow the approach described by
B\"oker et al.  (1999).  After re-binning, and after excluding a
user-specified percentage of the brightest pixels (due to, e.g.,
foreground stars), we plot the intensity, in counts per second, of
each pixel in the \ha\ image versus the intensity of the same pixel in
the continuum image. As described by B\"oker et al. (1999), in the
absence of line emission and in the case of constant colour across the
field, the resulting relation should be a straight line, down to the
noise level. Deviations from this straight line will be due to pixels
which trace colour variations due to different stellar populations,
differential extinction effects, \ha\ emission, or a combination of
these.  The plot will show most pixels in the image located along a
narrow band, with the minority of pixels that trace \ha\ emission
located above it (see Fig.~\ref{fig1} for an illustration). Points
below the band are scarce because very few pixels will trace either
excess continuum emission, or reduced line emission.  Such points
could have their origin in, for instance, \ha\ absorption in
foreground stars, saturation in one of the two images, or sporadic
events like cosmic rays.  The slope of a line fitted to the continuum
band will give the scaling factor, and the origin of the line will
give an additional estimate of the background in both the line and
continuum images (Fig.~\ref{fig1}).

While in a number of well-defined cases (high-quality \ha\ images,
and an accompanying continuum image taken through the 6470/115 or $R$
filter) the slope of the line as described above gives a very good
determination of the scaling factor for the continuum subtraction, we
checked the result of this first method in all cases using a second
method.  Here, we measure the integrated intensity of a number of
non-saturated foreground stars in the line and continuum images, and
deduce the scaling factor from the average ratio of these intensities.
Some stars have enhanced or depressed \ha\ emission, and in the
absence of further information on the stars imaged we identified such
stars by their deviant line/continuum intensity ratio.  This problem
was not severe because most of our \ha\ images were taken through \ha\
filters that were not sensitive to \ha\ emission at rest wavelength.

We found that the second method gave consistent results, both
internally (different stars in the same image) and externally (in
comparison to the pixel-to-pixel method described above). For example,
in NGC~3368, the scaling factor determined using the first method of
comparing the continuum and line counts at each pixel is 0.0535,
whereas the second method, using the foreground stars in the image,
gives a scaling factor of 0.0545. We used the foreground star method
in all cases to check and constrain the scaling factor as derived from
the pixel-to-pixel method.  In those cases where the pixel-to-pixel
method did not give conclusive results, a scaling factor as derived
from the foreground stars was used.

The minimum number of foreground stars needed to obtain a reasonable
constraint on the scaling factor was found to be six, a number which
results in a typical error of around 5\% in the scaling factor. With
more stars available, and/or with further constraints from the
pixel-to-pixel method, the uncertainty is even lower. Since the
$R$-band covers the \ha\ line, the flux in the \ha\ image was
corrected by about 3\% whenever an $R$ image was used as continuum
(cf. James et al. 2004). To produce the continuum subtracted image,
the continuum image was simply multiplied by the scaling factor, and
the result subtracted from the line image.  Scaling factors as used
are listed in Table~\ref{imagetab}.  Identical filter combinations can
give rather different scaling factors because the observing conditions
are not always the same for both images.  If, for instance, the
continuum image is observed on a night with higher extinction (due to,
e.g., thin clouds or to dust in the atmosphere), the scaling factor
for that image will be higher than if the conditions would have been
better.

We performed a number of specific tests to define the impact of the
uncertainties introduced by the scaling factor on the luminosities of
individual \hii\ regions, especially when using the $I$-band filter.
In both sets of tests, we compare integrated intensities of selected
well-defined \hii\ regions at different distances from the nucleus of
a galaxy.  Both changes in the scaling factor, and in the wavelength
of the continuum image, will translate into smaller errors further out
into the disk, where the level of continuum emission, as well as the
effects of dust, will be reduced as compared to the central regions.
In the first set of tests we compared \hii\ region luminosities as
measured from \ha\ images from which appropriately scaled 6470/115,
$R$- and $I$-band images had been subtracted.  We found that the
6470/115 and $R$ continua lead to random errors in \hii\ region
luminosities that are less than 2\%, without any systematic error. The
6470/115 and $I$ continua lead to similar random errors, but also to a
slight systematic difference of up to 5\%.  We find no evidence for an
increase in this error with decreasing distance to the nucleus.  In
the second set of tests, we compared the integrated intensities of
selected \hii\ regions while changing the scaling factor within the
range of uncertainty that accompanies our determination of that
factor.  Here we do find, not surprisingly, that uncertainties in the
\hii\ region luminosities are higher for those in the circumnuclear
regions (at about 10-15\% for the nucleus itself and the innermost
\hii\ regions, namely those within 0.5 kpc of the nucleus) than for
those in the disk (at about 2-5\%).  We conclude from these tests that
(1), although not ideal, it is certainly safe to use $I$-band images
for continuum subtraction, (2), the best results are obtained by
using either 6470/115 or $R$-band images, and (3), uncertainties
in the continuum subtraction lead to uncertainties of a few percent at
most in \ha\ luminosity and flux in the disks of our sample galaxies,
although higher in the nuclei and circumnuclear regions, where they
reach up to 10-15\%.

\subsection{Photometric calibration}

Many of our images were either taken under non-photometric conditions,
or (in the case of all images obtained from the archive) no guarantee
could be found that conditions during the observations had in fact
been photometric.  We combined all good images, from photometric and
non-photometric nights, and calibrated them once the final, combined,
images had been produced.  For this purpose, individual galaxy images,
as well as standard star exposures, obtained during those of our own
observing nights for which we could guarantee photometric conditions,
were maintained separately throughout the reduction process.  We used
a number of photometric nights during our final observing runs to
obtain short calibration images of those galaxies and in those filters
where up to that moment we had no calibration.  In the end all $B$,
$I$ and \ha\ images, and 44\% of the $R$-band images, have been
calibrated photometrically. 

All photometric images were calibrated using the images of standard
stars, mostly selected from Landolt's (1992) list for the broad band
observations, and from a variety of sources for the narrow-band
imaging (including Massey et al. 1988; Oke 1990).  Following the same
procedure as in Paper~I, these calibrated but short images were then
used to calibrate the images in the final data set.  This was done by
comparing the total flux in a certain area in the image to the flux as
measured in exactly the same area in the other image.  Normally, this
area was circular and centred on the galaxy's nucleus, except where
the nucleus was saturated in one of the images, in which case an
annular area was used. In two galaxies, emission from foreground stars
was used in addition to that from the galaxy itself. In all cases,
differences between the two images, for instance in pixel size and
background, were carefully taken into consideration. 

\begin{figure}
\psfig{figure=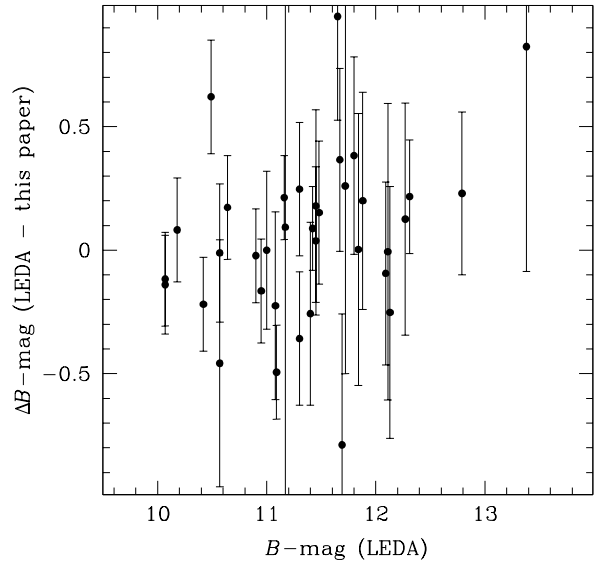,angle=0,width=8cm}
\caption{Comparison between the total $B$-magnitudes for our sample
galaxies as determined in this paper and as obtained from LEDA. The
error bars indicate a combination of uncertainties as derived for our
data and as given by LEDA, with the latter being the dominant source
of uncertainty.}
\label{phot_comparison}
\end{figure}

The resulting photometric calibration was checked in a variety of ways
against values published in the literature. For instance, we validated
that radial profiles derived from our images compare favourably with
the ones published by de Jong \& van der Kruit (1994): differences are
below 0.2~mag in the three galaxies which occur in both samples. We
also considered differences between the total magnitudes derived from
our imaging and those collected in the Lyon-Meudon extragalactic
database (LEDA). The results, shown in Fig.~\ref{phot_comparison}, are
similar to those reported by de Jong \& van der Kruit (1994, their
fig.~6), who compared total magnitudes as found by them and as
published in the RC3 (de Vaucouleurs et al. 1991). As in their case,
we find a large scatter, but crucially around a value of zero in the
magnitude difference. The scatter is due to a variety of reasons, such
as the data points compiled in LEDA, the difficulties in determining
total magnitudes from our data in cases where the imaged area of sky
is not much larger than the galaxy, or our own calibration.  Overall,
we are confident of the photometric calibration of our images, without
claiming high-precision photometry.

\section{Description of the final data set}

The final data set for each of our 57 sample galaxies consists of
images in the broad $B$ and $I$ bands, as well as in \ha.  In
addition, \ks\ images were obtained for all galaxies, as described in
detail in Paper~I.  For all except two galaxies (NGC~1300 and
NGC~7727), $R$-band images were obtained, and in some cases (NGC~628,
NGC~3184, NGC~3351, NGC~3631, NGC~4321, NGC~4395, NGC~5371, and
NGC~5457) we obtained also a $V$-band image.  All images were reduced
to equivalent exposure times of 1~sec, and were placed on the same
pixel grid for each galaxy.  This latter step implies that all images
in the final data set have pixels of 0.241~arcsec squared (the
smallest pixel size used, namely that of the near-IR INGRID camera,
see Paper~I), and that all images of any given galaxy are exactly
aligned to each other, are correctly oriented (to within the accuracy
dictated by the DSS images), and are of the same size.  For each
galaxy, we also produced a continuum-subtracted \ha\ image.

We updated the headers of the FITS image files to include a
description of the world coordinate system. For this, we used the
position of the galaxy as given on the NASA-IPAC Extragalactic
Database, NED, and tied this in with the centre position on our
images. We also included a keyword in the header of each image giving
the calibration constant which can be used to derive surface
brightnesses and magnitudes from the images. The images described in
this paper will be made available for general use through the CDS.

\section{Conclusions}

We present an extensive set of optical broad- ($B$, $R$, and $I$) and
narrow-band (\ha) images of a sample of 57 Northern, not highly
inclined, spiral galaxies of all types.  The images were obtained
partly from the literature and from the ING data archive, but mostly
from our own observations with the 1~m JKT and the 2.5~m INT on La
Palma.  In this paper, we describe the observations and the data
reduction procedures. Because of the characteristics of the data set,
obtained over the years with different instrumentation, we developed a
set of IRAF scripts, described in detail here, which allow us to
handle the volume of data involved in this project. We also describe
in some detail the procedures followed to subtract the continuum from
our \ha\ emission line images, and estimate uncertainties associated
with the continuum subtraction. For each of our 57 galaxies, our final
data set consists of aligned images in the $B, I$ and $K_{\rm s}$
(from Paper~I) broad bands, as well as in \ha. For 55 of these
galaxies, the set also contains an $R$-band image, while a $V$-band
image is available for eight galaxies.

We will use the image set in subsequent papers to investigate the
distribution and properties of the \hii\ regions in the disks and
central regions of the galaxies, the properties of their bars and
central regions, and the mutual influence of star formation and spiral
arms.  The images described in this paper will be made available for
general use through the CDS.

\begin{acknowledgements}

DMB and SLF thank the ING for the hospitality and financial support
they received during their stay as placement students.  Dr. Phil A.
James kindly let us use his \ha\ continuum filter.  Dr. Torsten
B\"oker is acknowledged for discussions about continuum subtraction.
The William Herschel, Isaac Newton and Jacobus Kapteyn Telescopes are
operated on the island of La Palma by the ING in the Spanish
Observatorio del Roque de los Muchachos of the Instituto de Astrof\'\i
sica de Canarias.  We thank the various service observers and service
programme managers who helped us obtain significant amounts of the
data presented here through the service programme.  Data were partly
retrieved from the ING archive.  The Digitized Sky Survey was produced
at the Space Telescope Science Institute under US Government grant NAG
W-2166. This research has made use of the LEDA database
(http://leda.univ-lyon1.fr), and of the NED. The latter is operated by
the Jet Propulsion Laboratory, California Institute of Technology,
under contract with the National Aeronautics and Space Administration.

\end{acknowledgements}


\begin{thebibliography}{}

\bibitem[B{\" o}ker et al.~1999]{1999ApJS..124...95B} B{\" o}ker, T., 
Calzetti, D., Sparks, W., et al. 1999, \apjs,  124, 95 

\bibitem[de Jong \& van der Kruit(1994)]{1994A&AS..106..451D} de Jong, 
R.~S.~\& van der Kruit, P.~C.\ 1994, \aaps, 106, 451 

\bibitem[\protect\citename{de Vaucouleurs} 1948]{1948AnAp...11..247D} de 
Vaucouleurs, G. 1948, AnAp,  11, 247

\bibitem[de Vaucouleurs et al.~1991]{1991RC3...C......0D} de Vaucouleurs 
G., de Vaucouleurs A., Corwin J.~R., Buta R.~J., Paturel G., Fouque P., 
1991, Third reference catalogue of Bright galaxies, 1991, New York : 
Springer-Verlag (RC3)

\bibitem[\protect\citename{Elmegreen} 1987]{1987ApJ...314....3E} Elmegreen, 
D.~M., Elmegreen, B.~G. 1987, ApJ, 314, 3

\bibitem[\protect\citename{Freeman} 1970]{1970ApJ...160..811F} Freeman, 
K.~C. 1970, ApJ,  160, 811

\bibitem[\protect\citename{Hubble} 1926]{1926ApJ....64..321H} Hubble, E.~P. 
1926, ApJ,  64, 321

\bibitem[James et al.(2004)]{2004A&A...414...23J} James, P.~A.~et al.\ 
2004, \aap, 414, 23 

\bibitem[\protect\citename{Jarrett} 2003]{2003AJ....125..525J}
Jarrett, T.~H., Chester, T., Cutri, R., Schneider, S.~E., \& Huchra,
J.~P. 2003, AJ, 125, 525

\bibitem[Jogee et al.(2002)]{2002ApJ...570L..55J} Jogee, S., Knapen, J.~H., 
Laine, S., Shlosman, I., Scoville, N.~Z., \& Englmaier, P.\ 2002, \apjl, 
570, L55

\bibitem[\protect\citename{Knapen} 1998]{1998MNRAS.297..255K} Knapen, J.~H. 
1998, MNRAS,  297, 255

\bibitem[Knapen et al.~1993]{1993AJ....106...56K} Knapen, J.~H., 
Arnth-Jensen, N., Cepa, J., \& Beckman, J.~E. 1993b, \aj, 106, 56

\bibitem[Knapen et al.(1993)]{1993ApJ...416..563K} Knapen, J.~H., Cepa, J., 
Beckman, J.~E., Soledad del Rio, M., \& Pedlar, A.\ 1993a, \apj, 416, 563 

\bibitem[Knapen, de Jong, Stedman, \& Bramich(2003)]{2003MNRAS.344..527K} 
Knapen, J.~H., de Jong, R.~S., Stedman, S., \& Bramich, D.~M.\ 2003, 
\mnras, 344, 527 (Erratum MNRAS, 346, 333; Paper~I)

\bibitem[\protect\citename{Landolt} 1992]{1992AJ....104..340L} Landolt,
A.~U. 1992, AJ,  104, 340

\bibitem[Massey, Strobel, Barnes, \& Anderson(1988)]{1988ApJ...328..315M} 
Massey, P., Strobel, K., Barnes, J.~V., \& Anderson, E.\ 1988, \apj, 328, 
315 

\bibitem[Oke(1990)]{1990AJ.....99.1621O} Oke, J.~B.\ 1990, \aj, 99, 1621 

\bibitem[\protect\citename{Rozas} 1996]{1996A&A...307..735R} Rozas, M., 
Beckman, J.~E., \& Knapen J.~H. 1996, A\&A,  307, 735

\end{thebibliography}
\end{document}